\documentclass[fleqn,usenatbib]{mnras}

\usepackage{newtxtext,newtxmath}

\usepackage[T1]{fontenc}

\DeclareRobustCommand{\VAN}[3]{#2}
\let\VANthebibliography\thebibliography
\def\thebibliography{\DeclareRobustCommand{\VAN}[3]{##3}\VANthebibliography}

\usepackage{graphicx}	
\usepackage{amsmath}	

\newcommand{\bri}{$(G_{\rm BP}-G_{\rm RP})_0$}

\newcommand{\mg}{$M_{G}$}

\title[Wave-like Oscillations in Mono-age Stars]{The wave-like disk oscillations of mono-age stellar populations in the Solar neighbourhood from Gaia DR3}

\author[Tao Wang et al.]{
Tao Wang,$^{1}$
Bing-Qiu Chen,$^{1}$\thanks{E-mail: bchen@ynu.edu.cn}
Jian-Hui Lian,$^{1}$
Mao-Sheng Xiang,$^{2}$
Xiao-Wei Liu$^{1}$
\\
$^{1}$South-Western Institute for Astronomy Research, Yunnan University, Kunming, 650500, P. R. China\\
$^{2}$National Astronomical Observatories, Chinese Academy of Sciences, Beijing, 100012, P. R. China
}

\date{Accepted XXX. Received YYY; in original form ZZZ}

\pubyear{2024}

\begin{document}
\label{firstpage}
\pagerange{\pageref{firstpage}--\pageref{lastpage}}
\maketitle

\begin{abstract}
The North-South asymmetry in the number density and bulk velocity of stars in the solar neighborhood provides valuable insights into the formation and evolution of the Milky Way disk. Our objective is to investigate the wave-like disk oscillations of mono-age stellar populations in the Solar neighbourhood using data from Gaia Data Release 3. We have selected a comprehensive sample of main sequence turn off stars. The ages of these stars can be accurately determined using isochrone fitting methods. Our findings indicate that the north-south density and mean vertical velocity asymmetries remain consistent across all age groups.The uniformity of perturbations across all subsamples suggests that all populations are responding to the same external influence, which likely affects them irrespective of their age. Moreover, the fact that these perturbations appear consistently implies they could be either ongoing or recent. Regarding vertical velocity dispersions, we observe that older stars exhibit larger dispersions.
\end{abstract}

\begin{keywords}
Galaxy: disc -- Galaxy: structure -- solar neighbourhood -- Hertzsprung-Russell and colour-magnitude diagrams -- Galaxy: kinematics and dynamics
\end{keywords}



\section{Introduction} \label{sec:intro}

In the Solar Neighbourhood, a North-South asymmetry in the vertical distribution of stellar number density within the Galactic disk has been observed {\citep{Widrow2012, Chen2017, Bennett2019}. This asymmetry is predominantly manifested as wave-like disturbances in both vertical stellar counts and vertical motions {\citep{Yanny2013,Sun2015,Bennett2019}. Recent research findings have indicated a potential connection between the observed asymmetry and the existence of a snail-like feature in the distribution of vertical phase-space, commonly known as the phase-space spiral {\citep{Antoja2018}. The phase-space spiral is interpreted as an incomplete mixing phenomenon occurring in the vertical direction of the Milky Way, exhibiting wave-like characteristics in phase-space diagrams {\citep{Bland2019,Laporte2019}. The existence of these asymmetric phenomena and the distinct snail-like features strongly imply the existence of intricate dynamical structures in the vertical direction near the Sun, potentially influenced by external perturbations or the transit of satellite galaxies {\citep{Li2020}. 

Investigating the aforementioned asymmetric phenomenon in various stellar populations characterized by different ages can yield vital insights into comprehending its underlying mechanisms and ramifications. These insights, in turn, offer valuable clues concerning the dynamics of the Milky Way. With the release of Gaia Data Release 3 (Gaia DR3; \citealt{GaiaDR32023}), a comprehensive dataset encompassing astrometry, multi-band photometry, and low-resolution spectroscopy for billions of stars has become available. Notably, Gaia DR3 presents us with the stellar ages of about 300 million sources ($G <18.25$) through the utilization of the Final Luminosity Age Mass Estimator (FLAME; \citep{Apsis2023}). These ages are determined by comparing the measured stellar luminosity and effective temperature with theoretical stellar evolution models. The efficacy of this method is particularly pronounced for the main sequence turn-off (MSTO) or subgiant stars, as stars at this specific evolutionary stage exhibits considerable variations in atmospheric parameters corresponding to their ages. 

This study aims to investigate the North-South asymmetry of the Galactic disk in the Solar neighborhood by analyzing the MSTO samples from DR3. Section 2 provides an introduction to the dataset, while Section 3 outlines the methodology employed. The obtained results are presented in Section 4. Furthermore, Section 5 offers a discussion on the observed asymmetry in vertical motions. Finally, Section 6 is a brief summary.

\begin{figure*}
\centering
\includegraphics[height=\columnwidth]{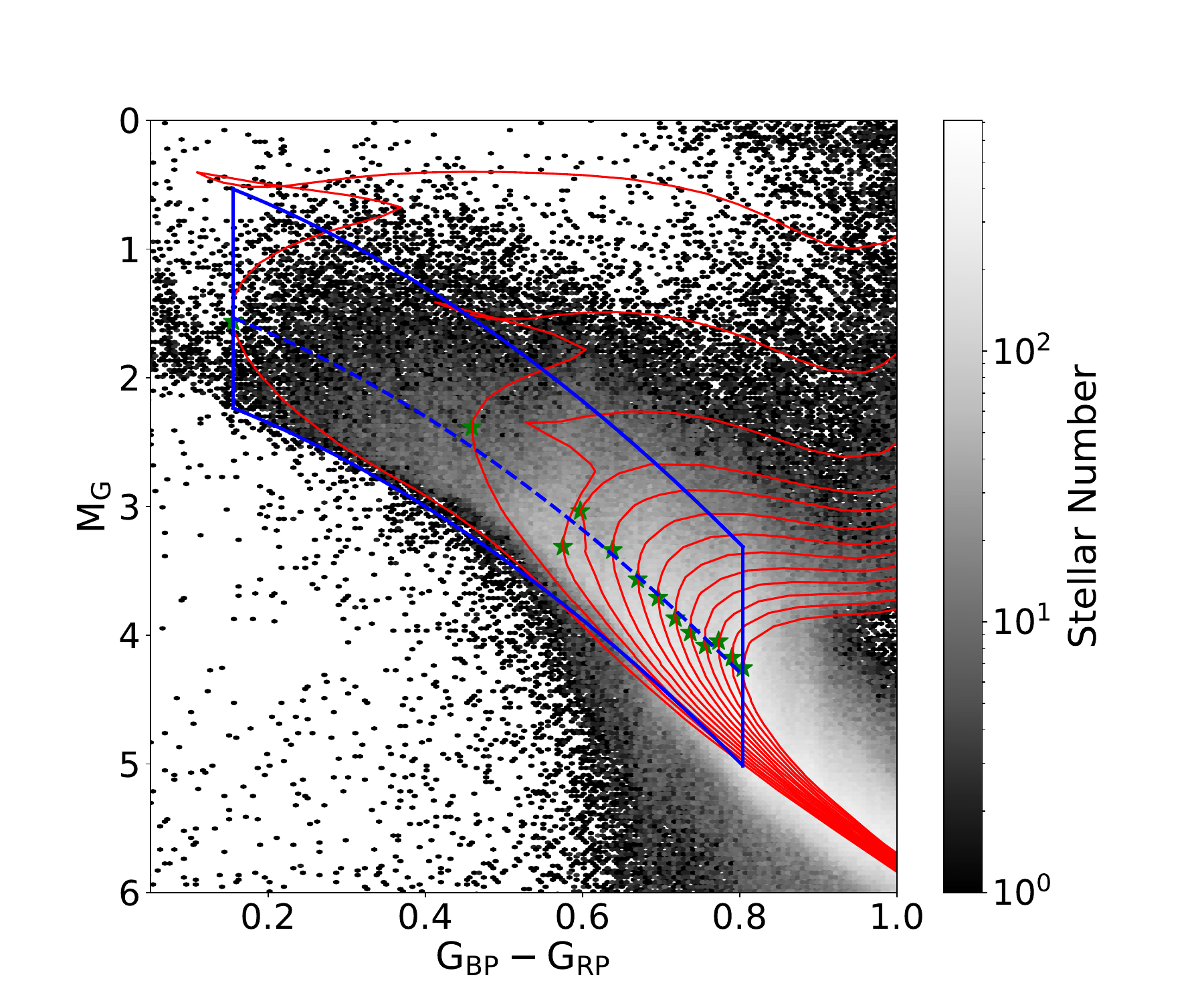}
\caption{The selection of the MSTO stars. The red solid lines correspond to the PARSEC isochrones, covering a range of ages from 1 to 13\,Gyr. The green star symbols indicate the precise positions of MSTO stars, where the isochrones start to deviate from the main sequence. The blue dashed line represents the fitting of these turning points. and the blue solid lines denote the ranges we consider when selecting MSTOs on the HR diagram.
\label{1}}
\end{figure*}

\section{Data}\label{sec:style}

This study utilizes data from Gaia DR3 \citep{GaiaDR32023}. To ensure a complete sub-sample, we have employed a simple magnitude cut of 7 $<G<$ 18.25\,mag. Furthermore, we have restricted our selection to stars within a cylindrical volume centered on the Sun with a height of 2\,kpc and a radius of 0.25\,kpc. This approach has yielded a sample of about 5.3 million stars. We have examined the $G$-band magnitude distribution of Gaia DR3 stars within a 0.25\,kpc radius from the Solar center. The magnitudes in this sample range from 7 to 21.5\,mag, with stellar counts decreasing beyond 20\,mag. This observation validates our decision to set the upper magnitude limit at 18.25\,mag, ensuring the sample's completeness within the selected range. In this study, we adopt the stellar extinction and distance informations provided by the General Stellar Parametriser for Photometry (GSP-Phot; \citealt{GSPph2023}). Additionally, we have assessed the impact of using alternative sources, such as the 3D extinction maps by \citet{Guo2021} and distance estimates by \citet{BailerJones2021}. Our analysis confirm that these modifications do not affect the selection of MSTO stars or alter the study's outcomes, thereby supporting the robustness of our conclusions against these updates in extinction and distance estimates.

We then define a sample of MSTO stars, which are selected based on their intrinsic colour \bri\ and absolute magnitude \mg\ around the main sequence turn-off. The intrinsic color of these stars is highly sensitive to their ages, making them ideal candidates for age estimation using the technique of stellar isochrone fitting. To identify the MSTO stars, we define ranges of \bri\ and \mg\ for different ages of MSTO stars using Padova stellar isochrones \citep{Bressan2012} with [Fe/H] = 0\,dex. As depicted in Fig.~\ref{1}, we trace the trajectories of these stars in the HR diagram. After mapping the colours and magnitudes of MSTO stars across various ages, we have fitted a curve through these data points and adjusted it by shifting 1\,mag upward in $M_G$ and 0.7\,mag downward within the defined range (blue box in Fig.~\ref{1}). All stars falling within this range are considered MSTO stars. This yields as sample of 347,715 stars.

We adopt the age estimates for the selected MSTO stars from the Final Luminosity Age Mass Estimator (FLAME; \citealt{Apsis2023}) in Gaia DR3. FLAME derives stellar parameters, including the radius, luminosity, gravitational redshift, mass, age, evolutionary stage, and auxiliary parameters from the outputs of the GSP-Phot and  the General Stellar Parametriser for Spectroscopy (GSP-Spec; \citealt{GSPsp2023}), along with astrometry, photometry and stellar models. The errors in the ages of these stars are mostly below 25\%.  The median value of the relative errors is 15.8$\%$.  We cross-match our sample with the LAMOST MSTO sample from \citet{Xiang2017} and obtain a total of 14,819 common stars. In Fig.~\ref{2}, the left panel shows a comparison between the ages derived from FLAME and those reported by \citet{Xiang2017}, demonstrating general agreement, though with some systematic discrepancies for stars older than $\sim$10 Gyr, where the ages from \citet{Xiang2017} are systematically higher. To further validate the FLAME age estimates, we have compared them with asteroseismic ages from \citet{Silva2017} using Kepler data (right panel of Fig.~\ref{2}). Despite the small sample size of 43 common stars, the correlation between FLAME and asteroseismology ages is strong, with age differences remaining within acceptable limits, although FLAME ages are slightly overestimated.
    
\begin{figure*}
    \centering
    \includegraphics[height=0.8\columnwidth]{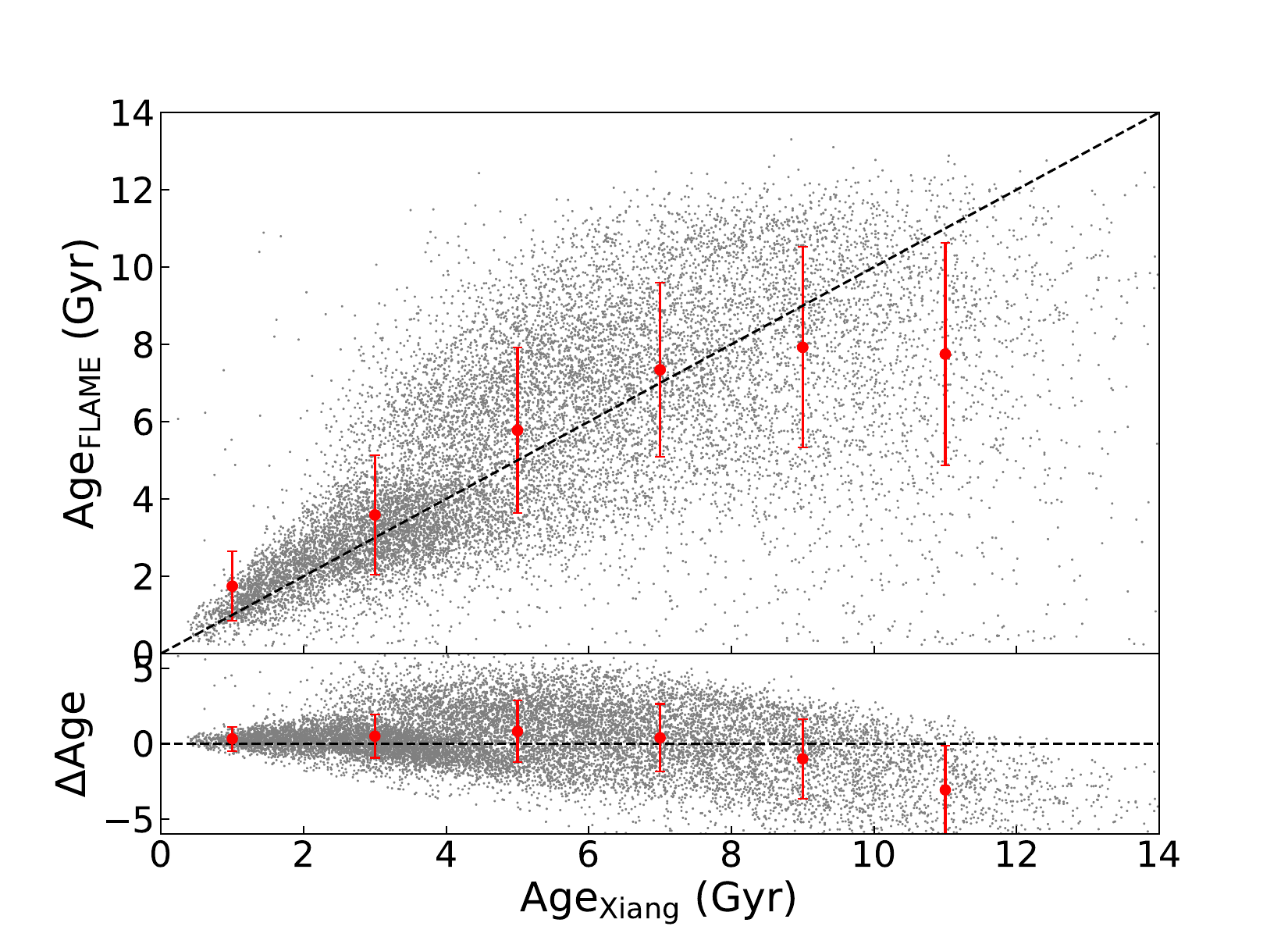}
    \includegraphics[height=0.8\columnwidth]{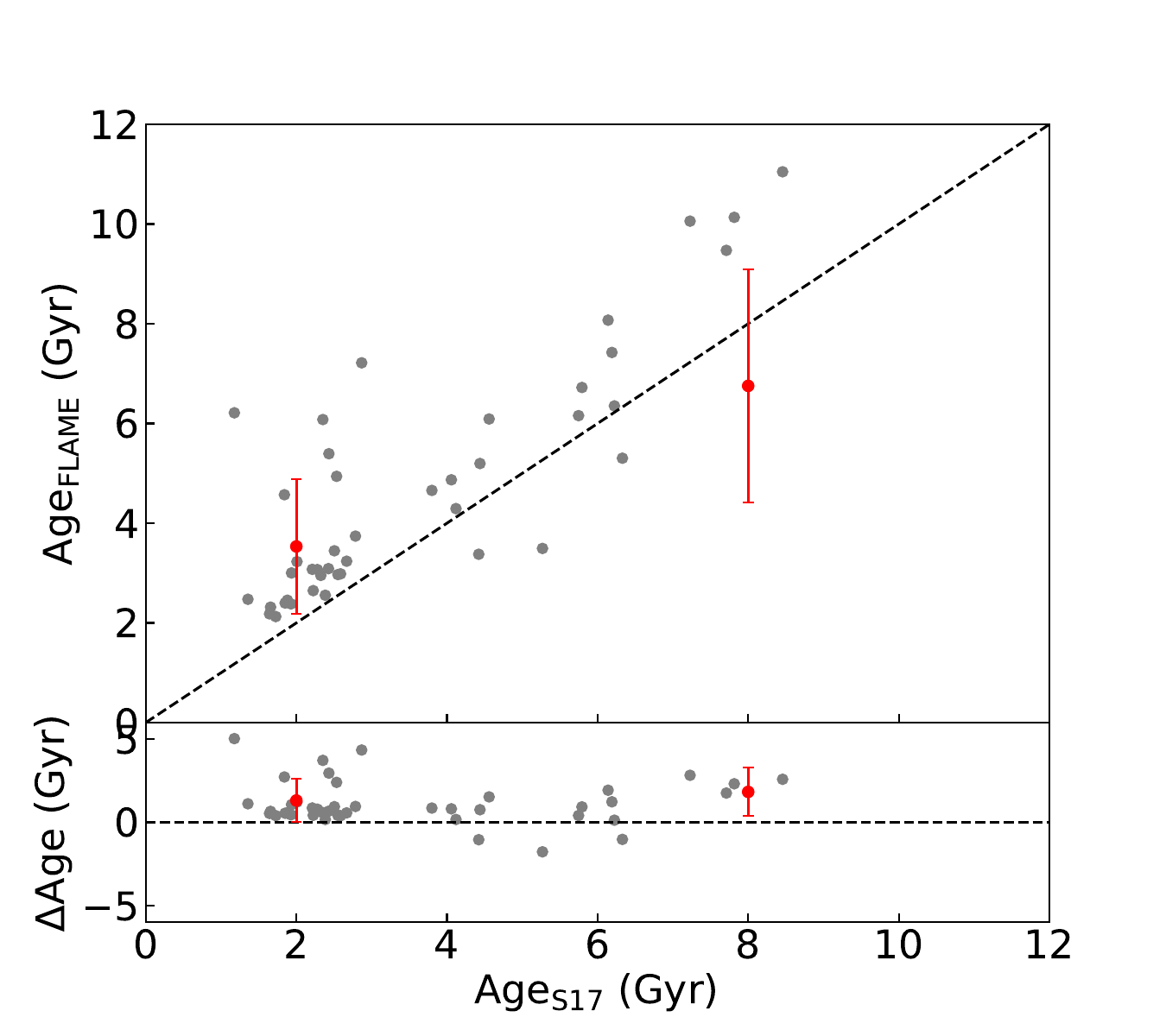}
    \caption{Comparative Analysis of Age Estimates. \textit{Left panel}: Comparison of FLAME ages and those obtained from \citet{Xiang2017} for 14,819 overlapping stars. The red points with error bars indicate the mean values and their dispersion within each age bin. \textit{Right panel}: Comparison between FLAME ages and asteroseismic ages from \citet{Silva2017} for 43 common stars, also marked by red points indicating mean and dispersion within each age bin.}
    \label{2}
\end{figure*}

\begin{figure*}
\centering
\includegraphics[width=0.8\textwidth]{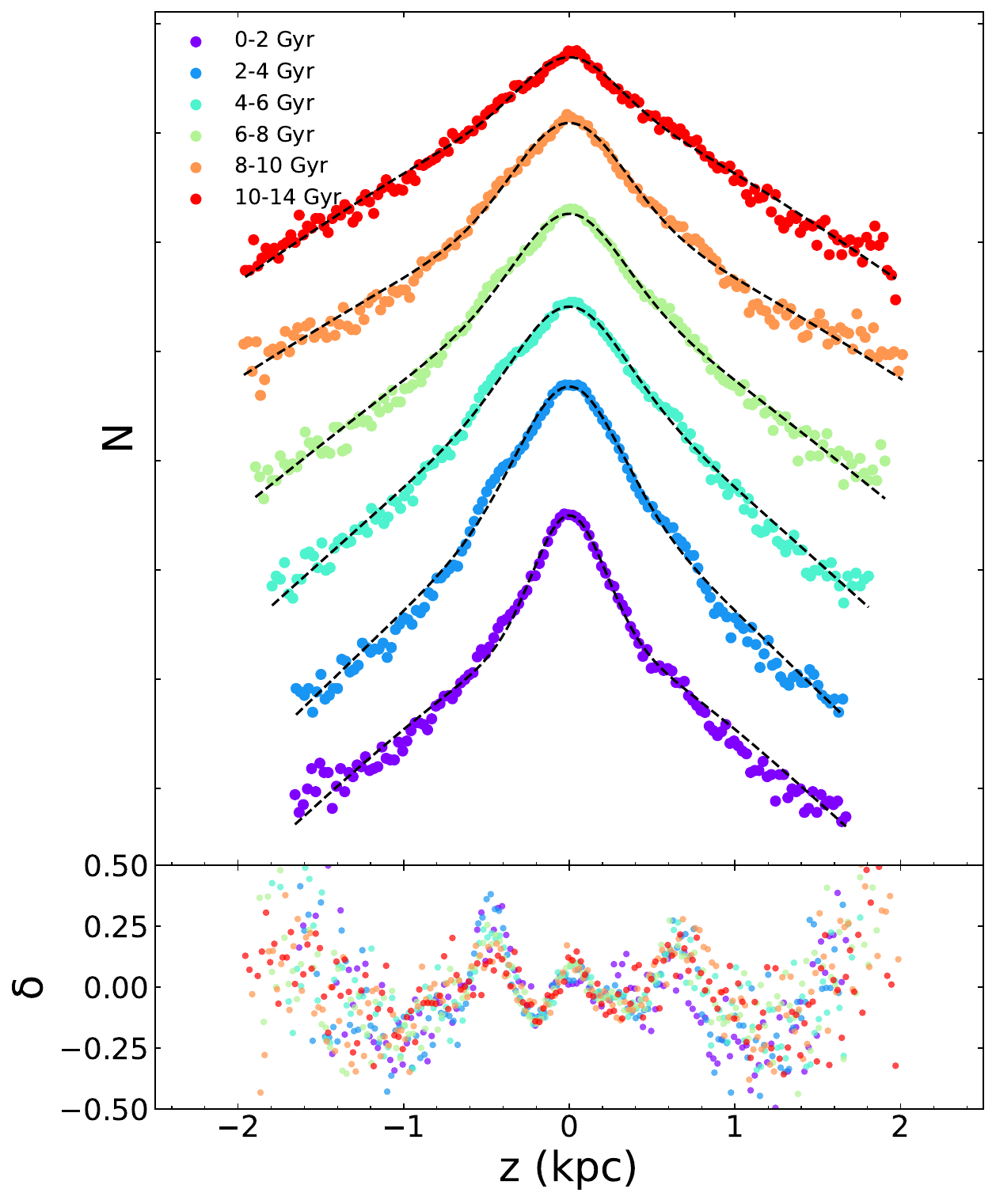}
\caption{Best-fit number density profiles (upper panel) and the residuals (bottom panels) for the individual mono-age stellar populations. The measured stellar number counts for various stellar populations are represented by dots of different colors in the plot. To improve visibility, the raw number counts have been vertically shifted along the y-axis. The dashed lines correspond to the best-fit models. 
\label{3}}
\end{figure*}

\section{Method} \label{sec:floats}

In this study, we have divided our sample of MSTO stars into mono-age populations based on their ages, and subsequently quantified their number counts ($N$) as a function of their distances from the Galactic plane ($z$) within each age bin. To fit the $N$-$z$ profile of each mono-age population, we have employed a simple disk model comprising of two distinct stellar components. Similar to that employed by \citet{Widrow2012}, the number count profiles of both components are characterized by sech$^2$ functions, given by,
\begin{equation}
N(z_{\text{obs}}) = N_0 \left( \text{sech}^2 \left( \frac{z_{\text{obs}} + z_{\sun}}{2H_1} \right) + f \text{sech}^2 \left( \frac{z_{\text{obs}} + z_{\sun}}{2H_2} \right)\right),
\end{equation}
where $N(z_{\text{obs}})$ represents the number counts at the position $z_{\text{obs}}$, $N_0$ denotes the number count of the first component at $z = 0\,\mathrm{pc}$, and $f$ represents the number ratio between the second component and the first component at the same position. $H_1$ and $H_2$ are the scale heights of the two components, while $z_{\sun}$ is the position of the mid-plane of the Galactic disk. We have also evaluated a three-component model for the structure of the Galactic disk, following the hypothesis of \citet{Binney2023} that the disk could be a superposition of multiple components. Our analysis, however, indicates that the three-component model does not provide a better fit than the two-component model. The residuals between these models are similar, suggesting that the additional complexity of a three-component model does not significantly enhance the model accuracy. Consequently, we have opted to continue with the simpler two-component framework for further analysis.

Given the nature of our count data, which does not adhere to a Gaussian distribution, we adopt a Poisson distribution for the likelihood calculations. The likelihood function is expressed as follows \citep{Bennett2019}:
\begin{equation}
\ln p\left(N_{\text{obs}} | N \right) = \sum_i \left[ -N_i + N_{\text{obs},i} \ln\left(N_i \right) - \ln\left(N_{\text{obs},i}! \right) \right]
\end{equation}
Here, \(N_i\) represents \(N(z_i)\) from equation (1), \(N_{\text{obs},i}\) is the observed count \(N_{\text{obs}}(z_i)\), and \(z_i\) denotes the midpoint of different vertical position intervals. The factorial term, being constant for a given dataset, does not impact the relative likelihoods and can be omitted in maximum likelihood estimation.}

For parameter estimation, we assume uniform priors over plausible ranges for each parameter associated with single-age populations, namely \(N_0\), \(f\), \(z_{\sun}\), \(H_1\), and \(H_2\). We utilize the Markov Chain Monte Carlo (MCMC) method for optimization, implemented via the \textit{emcee} package \citep{emcee}. This approach enables us to systematically explore the parameter space and identify the most likely parameter values.

To address the uncertainties associated with photometric, parallax, extinction, and age measurements, we employ a Monte Carlo (MC) simulation approach. By generating 100 synthetic datasets that incorporate these uncertainties, we analyze each dataset using our fitting algorithm. The standard deviations of the derived parameters from these simulations are then used as estimates of their uncertainties.

\section{Results} \label{sec:floats}

We have fitted the the number count profiles for stellar populations in age bins of 0--2, 2--4, 4--6, 6--8, 8--10, and 10--14\,Gyr. Fig.~\ref{3} show the fitting results of all the populations, and the best-fit parameters are listed in Table~\ref{tab1}. Overall, the Eq.(1) captures the general features of the measured $N(z_{\text{obs}})$ for all populations. However, strong oscillations, i.e. the North–South asymmetry, features are visible for all populations, which will be discussed in Sect.~\ref{sec:Osc}. Table~\ref{tab1} shows that as the age increases, the scale heights of the disks increases, which is consistent with precious works (e.g., \citealt{Bovy2013,Xiang2018}). 

\begin{table*}
\caption{The best-fit parameters from the MCMC fits}
\begin{tabular}{lcccccc}
\hline
Age       & $N_0$ & $z_{\sun}$ (pc) & $f$       & $H_1$  (pc)   & $H_2$ (pc)    & \textbf{$N_{\rm obs}$}\\ 
\hline
0-2 Gyr   & $2714.94^{+15.79}_{-16.35}$ & $6.79^{+0.85}_{-0.85}$ & $0.10^{+0.00}_{-0.00}$ & $9.07^{+0.86}_{-0.89}$ & $357.32^{+6.35}_{-6.91}$ & 55367\\
2-4 Gyr   & $3906.69^{+27.28}_{-28.28}$ & $13.29^{+0.83}_{-0.84}$ & $0.09^{+0.01}_{-0.01}$ & $12.06^{+1.50}_{-1.61}$ & $336.12^{+6.54}_{-7.68}$ & 94966\\
4-6 Gyr   & $2119.56^{+31.96}_{-30.39}$ & $18.42^{+1.19}_{-1.20}$ & $0.18^{+0.03}_{-0.03}$ & $14.19^{+3.08}_{-3.25}$ & $362.79^{+9.27}_{-11.18}$ & 69941\\
6-8 Gyr   & $1558.93^{+20.68}_{-20.46}$ & $18.02^{+1.42}_{-1.42}$ & $0.18^{+0.02}_{-0.02}$ & $14.22^{+3.07}_{-3.26}$ & $403.80^{+8.95}_{-10.57}$ & 54140\\
8-10 Gyr  & $1146.93^{+12.26}_{-12.06}$ & $23.35^{+1.76}_{-1.76}$ & $0.15^{+0.02}_{-0.02}$ & $14.79^{+3.00}_{-3.11}$ & $501.71^{+10.64}_{-12.04}$ & 40808\\
10-14 Gyr & $410.20^{+13.16}_{-14.64}$ & $31.90^{+2.92}_{-2.93}$ & $0.45^{+0.07}_{-0.07}$ & $14.94^{+8.42}_{-9.27}$ & $465.29^{+12.38}_{-15.57}$ & 23350\\ 
\hline
\end{tabular}
\label{tab1}
\end{table*}

\subsection{Oscillations for the mono-age populations}
\label{sec:Osc}

In the bottom panel of Fig.~\ref{3}, we present the normorisized residuals, $\delta(z) = (N - \text{model}) / \text{model}$, for the best-fit models of the individual mono-age stellar populations. Overall, a distinct fluctuating pattern is evident. Starting from $z=0$, we observe the first prominent peak at vertical distances of approximately $z=0.1$\,kpc. The residuals at this peak are relatively small, of around $\delta=0.1$. As we move further away from the Galactic plane, two clear troughs become apparent, located at approximately $z = -0.2$\,kpc and $z = 0.5$\,kpc, respectively. The residuals at these positions are approximately $\delta=-0.2$. At greater distances, two additional peaks are observed, occurring at approximately $z = -0.5$\,kpc and $z = 0.7$\,kpc. The residuals at these positions are approximately $\delta=0.25$. Beyond 1\,kpc, the oscillations persist. Despite the presence of significant dispersion, the variation trend of $\delta(z)$ remains distinguishable.

The overall trend of variation observed in our study is consistent with previous works \citep{Widrow2012, Bennett2019}. A key finding of this research is the persistent identification of wave-like patterns in the residuals for all age groups. Unlike \citet{Bennett2019}, who noted similar patterns within samples divided by intrinsic color (an indirect proxy for stellar age), our analysis demonstrates that these patterns prevail across a more precisely defined age range.

Particularly, within the positional interval where Gaia distance measurements are most reliable ($|z| < 0.8\,\mathrm{kpc}$), the residuals for different age populations show remarkably little dispersion, suggesting a coherent response to vertical perturbations across these groups. This uniformity implies that the perturbative mechanism likely affects all stellar populations similarly, irrespective of their age.

This observation suggests that the perturbation is either currently active or occurred recently, as a historical event would have imprinted diverse effects on the stellar populations over time. Nonetheless, the transient nature of this perturbation remains unconfirmed due to the absence of long-term evidence. Detailed dynamical modeling is required to ascertain the specifics of this phenomenon. Although our results indicate a uniform perturbation impacting all age groups, they do not confirm its persistence over extended periods.

Moreover, the amplitude of fluctuations in the fit residuals, as shown in Fig.~\ref{3} (lower panel), diminishes for older stellar populations, aligning with expectations for dynamically hotter groups that have undergone more mixing and settling. In contrast, dynamically colder populations, being less mixed, typically exhibit more pronounced perturbations. Emphasizing this theoretical expectation provides deeper insights into the observed perturbations within the Galactic disk.

Our study also reveals an age-dependent variation in the vertical position of the Galactic mid-plane ($z_{\sun}$), ranging from 6.79 pc for the youngest stars to 31.9 pc for the oldest stars. These findings might elucidate discrepancies in earlier studies concerning the solar position, which utilized tracers from different age groups \citep{Binney1997, Chen2001, Juric2008}. However, the uncertainties associated with measurements of $z_{\sun}$ are comparable to these variations, suggesting the potential observational artifact nature of the age dependency of the Galactic mid-plane's position.

\section{Vertical velocities and dispersions}

\begin{figure}
\centering
\includegraphics[width=\columnwidth]{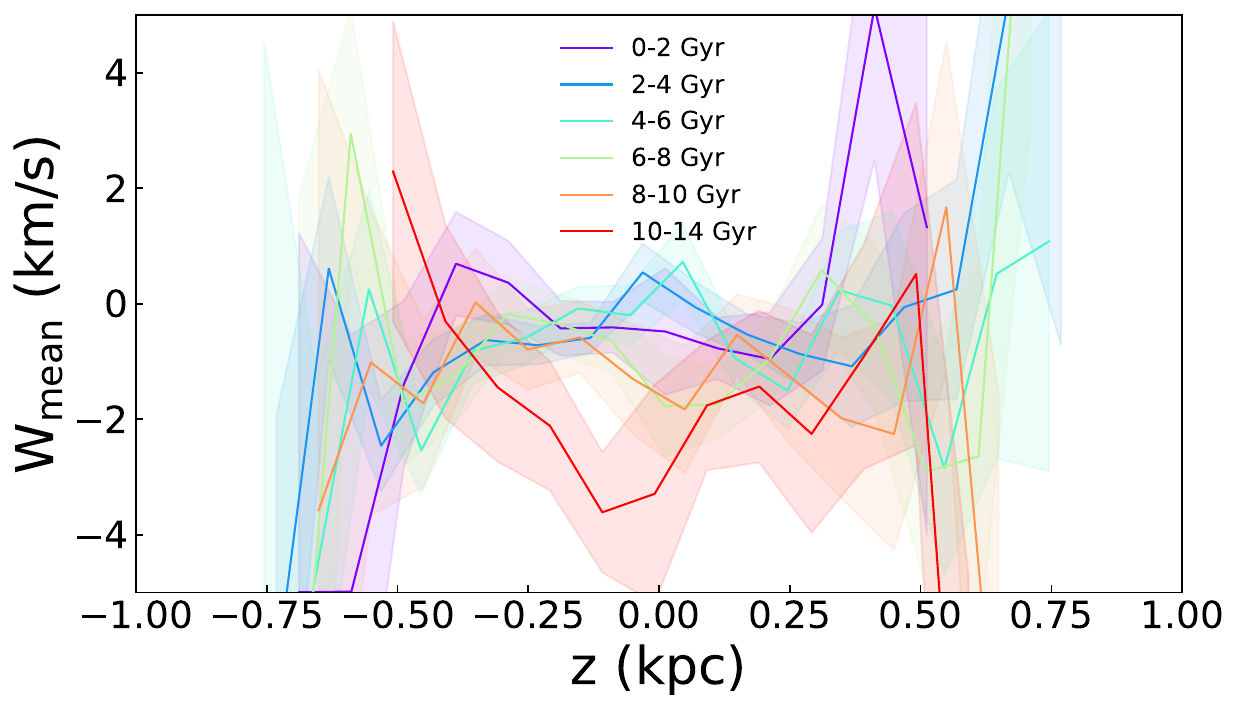}
\includegraphics[width=\columnwidth]{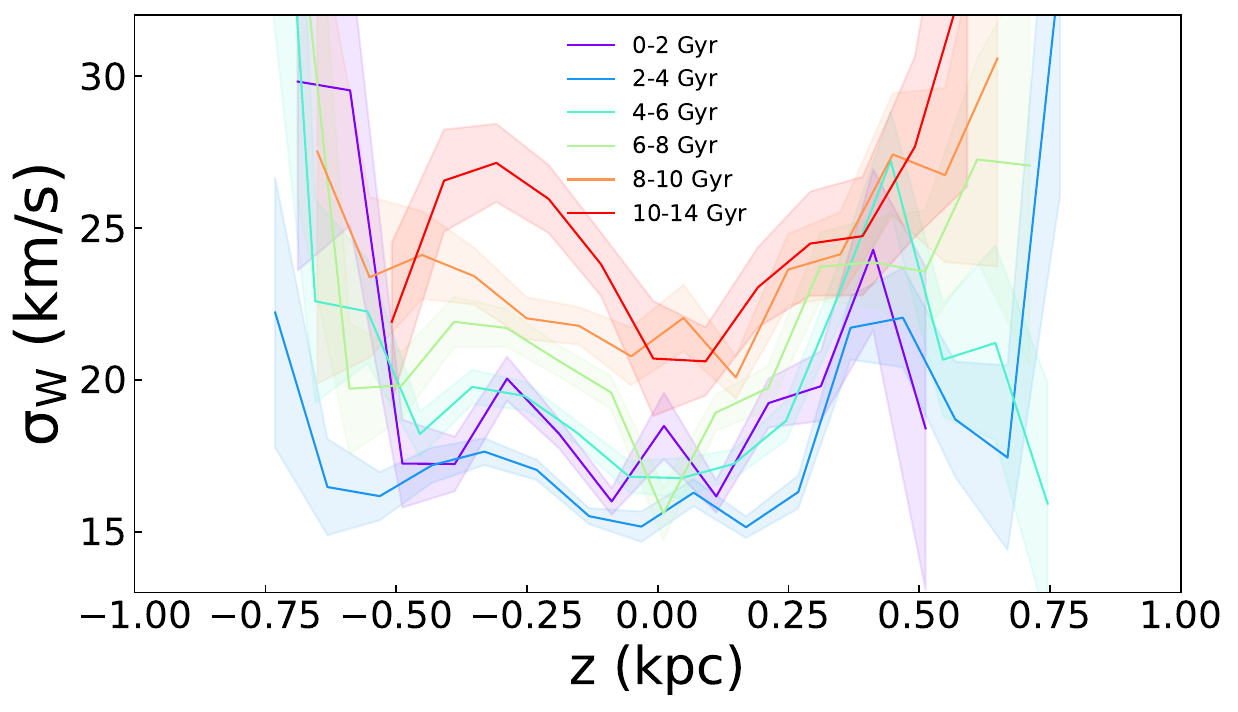}
\caption{ The mean (upper panel) and dispersion (bottom panel) values of the vertical velocities of stars in the individual stellar populations of different age ranges. Different stellar populations are represented by different colors. The typical uncertainties are labelled in the plots.
\label{4}}
\end{figure}

In addition to analyzing star counts statistically, we have investigated the kinematic properties of our samples, with a specific focus on the vertical velocity. Our sample comprises 49,550 stars with measurements of both their radial velocity and proper motions obtained from Gaia DR3. We calculated the vertical velocities ($W$) for these stars. In this study, we adopt the vertical velocity of the Sun as $W_{\sun} = 7.25\, \text{km}\,\text{s}^{-1}$ \citep{Sch2010}. Fig.~\ref{4} presents the mean and dispersion values of the derived vertical velocities for stars within each mono-age stellar population. We have used age bins identical to those employed for number density fitting, spanning from 0--2\,Gyr to 10--14\,Gyr, and performed calculations by binning every 100\,pc in the vertical direction.

In the analysis of the mean vertical velocity $W$ presented in the upper panel of Fig.~\ref{4}, notable variations are observed for $|z| > 0.5\,\text{kpc}$, but these are accompanied by significant uncertainties. Across the examined range of vertical distances, all stellar populations, segmented into various age bins, exhibit consistent trends. Specifically, reduced vertical velocities are noted at $z = 0.2\,\text{kpc}$ and $z = -0.22\,\text{kpc}$, correlating with peaks in stellar number density. These findings align with previous studies \citep{Bennett2019}. However, it is crucial to acknowledge that the trends in the upper panel are significantly influenced by noise, predominantly at greater heights above the galactic plane and notably in the 10--14 Gyr age cohort, likely due to sparse data.

The lower panel of Fig.~\ref{4} explores the variation in vertical velocity dispersion $\sigma_{W}$ as a function of $z$ position. The data indicate an increase in $\sigma_{W}$ with stellar age, reaching a maximum dispersion of $28\,\text{km s}^{-1}$ in the 10--14\,Gyr age bin. This trend of increasing dispersion with age is consistent across different stellar cohorts, underscoring the relationship between age and dynamic heating in the Galactic disk.

\section{Conclusions } \label{sec:style}

In our study, we have conducted a comprehensive analysis of MSTO stars selected from Gaia DR3. Our primary focus is to investigate the asymmetry between the northern and southern regions in groups of stars with different ages within the solar neighborhood of the Milky Way disk. Our analysis has revealed a persistent pattern of asymmetry in the vertical density across all age groups, suggesting the influence of a continuous or active mechanism.
    
A prominent discovery is the consistent wave-like pattern of vertical oscillations affecting stars of all ages, indicating that the observed phenomena result from ongoing or recent perturbations, rather than being relics of past events. In discussing the formation of this structure, we considered several possible scenarios. These include accretion due to the disruption of satellites, minor merger events, radial migration, and in-situ star formation during the process of gas accretion. Each of these factors significantly contributes to the observed vertical dispersion within the disk. These comprehensive considerations, including the use of more precise data and a variety of theoretical models, are essential for a deeper understanding of the complex mechanisms shaping the Galactic disk.

\section*{Acknowledgements}

This work is partially supported by the National Key Research and Development Program of China no. 2019YFA0405500, National Natural Science Foundation of China 12173034, 11803029, and 12322304 and Yunnan University grant No.~C619300A034. We acknowledge the science research grantsfrom the China Manned Space Project with No. CMS-CSST-2021-A09, CMS-CSST-2021-A08, and CMS-CSST-2021-B03. B.Q.C. acknowledges the National Natural Science Foundation of Yunnan Province 202301AV070002 and the Xingdian talent support program of Yunnan Province. JL acknowledges support by Yunnan Province Science and Technology Department under Grant No. 202105AE160021 and No. 202005AB160002 and by Yunnan University grant No. CY22623101.

This work presents results from the European Space Agency (ESA) space mission Gaia. Gaia data are being processed by the Gaia Data Processing and Analysis Consortium (DPAC). Funding for the DPAC is provided by national institutions, in particular the institutions participating in the Gaia MultiLateral Agreement(MLA). The Gaia mission website is https://www.cosmos.esa.int/gaia. The Gaia archive website is https://archives.esac.esa.int/gaia.

\section*{Data Availability}
Derived data supporting the findings of this study
are available from the corresponding author on request.



\bibliographystyle{mnras}
\bibliography{mnras_submit} 







\bsp	
\label{lastpage}
\end{document}